\documentstyle[12pt,epsf,aasms4]{article}

\def\gsim{\mathrel{\rlap {\raise.5ex\hbox{$ > $}}
{\lower.5ex\hbox{$\sim$}}}}
\def\lsim{\mathrel{\rlap {\raise.5ex\hbox{$ < $}}
{\lower.5ex\hbox{$\sim$}}}}

\newcommand{\be}{\begin{equation}}
\newcommand{\ee}{\end{equation}}
\newcommand{\bea}{\begin{eqnarray}}
\newcommand{\nn}{\nonumber}
\newcommand{\eea}{\end{eqnarray}}

\newcommand{\nk}{\noindent}

\def\gappeq{\mathrel{\rlap {\raise.5ex\hbox{$>$}}
{\lower.5ex\hbox{$\sim$}}}}
 
\def\lappeq{\mathrel{\rlap{\raise.5ex\hbox{$<$}}
{\lower.5ex\hbox{$\sim$}}}}
 
\begin{document}
\begin{titlepage}

\begin{flushright} 
CERN.TH/99--224 \\
ACT-7-99 \\
CTP-TAMU-30-99 \\
OUTP--99--35P \\
astro-ph/9907340
\end{flushright}

\begin{centering} 

{{\bf  Astrophysical Probes of the Constancy of the Velocity of Light}}

\vspace{0.05in} 

{\bf John Ellis} \\
{Theory Division, CERN, CH-1211 Geneva 23, Switzerland}

{\bf K. Farakos} \\
{Department of Physics, National 
Technical University of Athens, 
Zografou Campus, GR~157~80~Athens, Greece}

{\bf N.E. Mavromatos}\\
{Department of Physics, University of Oxford, 
1 Keble Road,  
Oxford OX1 3NP, U.K., and \\Theory Division, CERN, 
CH-1211 Geneva 23, Switzerland} 

{\bf V.A. Mitsou}\\
{Experimental Physics Division, CERN, 
CH-1211 Geneva 23, Switzerland. and \\ 
Department of Physics, University of Athens, 
Panepistimioupolis Zografou GR~157~71~Athens, Greece}

{\bf D.V. Nanopoulos} \\
{Department of Physics, 
Texas A \& M University, College Station, 
TX~77843-4242, USA,
Astroparticle Physics Group, Houston
Advanced Research Center (HARC), Mitchell Campus,
Woodlands, TX 77381, USA, and 
Academy of Athens, Chair of Theoretical Physics, 
Division of Natural Sciences, 28~Panepistimiou Avenue, 
GR~10679~Athens, Greece}
%

\newpage

{\bf Abstract}
\end{centering} 

{\small We discuss possible tests of the constancy of the
velocity of light using distant astrophysical
sources such as gamma-ray bursters (GRBs), Active
Galactic Nuclei (AGNs) and pulsars. This speculative
quest may be motivated by some models of quantum
fluctuations in the space-time background, and we
discuss explicitly how an energy-dependent 
variation in photon velocity
$\delta c/ c \sim  - E / M$
arises in one particular quantum-gravitational model.
We then discuss how data on GRBs may be used to set
limits on variations in the velocity of light, which
we illustrate using BATSE and OSSE observations of the GRBs
that have recently been identified optically and for which
precise redshifts are available. We show how a regression
analysis can be performed to look for an energy-dependent
effect that should correlate with redshift. The present
data yield a limit
$M \gsim 10^{15}$ GeV for the 
quantum gravity scale. We discuss the prospects for
improving this analysis using future data, and how one
might hope to distinguish any positive signal from
astrophysical effects associated with the sources.
}

\end{titlepage} 

\section{Introduction}

The constancy of the velocity of light is one of the most basic  tenets of
modern physics. It rests on a firm experimental basis and is embedded in
Lorentz invariance and the Special and General Theories of Relativity, as well
as Quantum Field Theory. It may therefore seem absurd to question the 
constancy of the velocity of light, and unnecessary to propose testing it. 
Nevertheless, we think
one should keep such a basic precept under review, and seize any new
experimental opportunity to test it more severely than before, particularly if
there are any theoretical speculations that might lead one to question it.

We believe~(\cite{nature}) 
that just such an opportunity to test the constancy of the velocity
of light is provided by transient astrophysical sources of $\gamma$ 
rays, such
as gamma-ray bursters (GRBs)~(\cite{Piran,Rees}), active galactic nuclei
(AGNs) and pulsars. We
also believe~(\cite{emnnew}) 
that modern approaches to the quantization of gravity provide some
motivation for exploring the opportunities provided by these astrophysical
sources. The primary purpose of this paper is to review these 
opportunities
and
to make some tentative steps towards exploiting them, by proposing and
pioneering some techniques  for analyzing the data. This is particularly timely
because we expect large amounts of useful observations of GRBs and AGNs to
become available in the near future.

The primary tool for measuring small variations $\delta c$ in the velocity
of light $c$ is the variation in arrival time~(\cite{aemn,nature}) 
$\delta t \simeq - (L/c)~(\delta c/c)$ observed
for a photon travelling a distance $L$. This clearly places a premium on
observing sources whose emissions exhibit structure on short time scales
$\Delta t \lappeq \delta t$ located at large distances. Some typical numbers
for some astrophysical sources are shown in Table 1. The relevant
photon property that could be correlated with velocity variations $\delta c$ is
its frequency $\nu$, or equivalently its energy $E$, for which characteristic
values are also listed in Table 1. As we discuss in more detail later, any such
effect could be expected to increase with $E$, and the simplest possibility,
for which there is some theoretical support, is that $\delta c \propto E/M$,
where $M$ is some high energy scale. In this case, the relevant figure of merit
for observational tests is the combination
\be
{L \cdot E \over c\cdot \delta t}
\label{one}
\ee
which measures directly the experimental sensitivity to such a high energy
scale M.
This sensitivity is also listed in Table 1,
where we see that our
favoured astrophysical sources are potentially sensitive to $M$ approaching
$M_P \simeq 10^{19}$ GeV, the mass scale at which gravity becomes strong.
Therefore, these astrophysical sources may begin to challenge any theory of
quantum gravity that predicts such a linear dependence of $\delta c$ on $E$.
Alternatively, it could be that $\delta c / c \simeq (E/\tilde M)^2$, in which
case the appropriate figure of merit is $E\sqrt{L/c \delta t}$, which is also
listed in Table 1. In this case, we see that astrophysical observations may be
sensitive to $\tilde M~\sim~10^{11}$ TeV. 

{\small 
\begin{table}[ht]
\begin{center}
\begin{minipage}{\linewidth}
\renewcommand{\thefootnote}{\fnsymbol{footnote}}
\begin{center}
{\bf Table~1:~Observational~Sensitivities~and~Limits~on~$M,{\tilde M}$}
\end{center}
\begin{tabular}{|c||c|c|c|c|c|}   \hline
Source & Distance & $E$ & $\Delta t$ & 
\parbox{3cm}{Sensitivity to $M$} &
\parbox{3cm}{Sensitivity to $\tilde M$}
\\ \hline
GRB 920229~\footnote{\cite{nature}, {\it see also}~\cite{schaefer}} & 3000
Mpc (?) & 200 keV & $10^{-2}$
s &
$0.6\times 10^{16}$ GeV (?) & $10^6$~GeV (?)
\\ \hline
GRB 980425~$^a$ & 40 Mpc & 1.8 MeV & $10^{-3}$ s (?)
& 
$0.7\times 10^{16}$ GeV (?) & $3.6 \times 10^6$~GeV (?) 
\\ \hline
GRB 920925c~$^a$ & 40 Mpc (?) & 200 TeV (?) & 200 s & 
$0.4\times 10^{19}$ GeV (?) & $8.9 \times 10^{11}$~GeV (?) 
\\ \hline
Mrk 421~\footnote{\cite{biller}} & 100 Mpc & 2 TeV & 280 s & 
$> 7 \times 10^{16}$ GeV & $> 1.2 \times 10^{10}$~GeV
\\ \hline
Crab pulsar~\footnote{\cite{crab}} & 2.2 kpc & 2 GeV & 0.35 ms & $> 1.3 \times
10^{15}$ GeV & $> 5 \times 10^7$~GeV
\\ \hline
GRB 990123 & 5000 Mpc & 4 MeV & 1 s (?) & $2 \times 10^{15}$ GeV (?)
& $2.8 \times 10^6$~GeV (?)
\\ \hline
\end{tabular} 
\end{minipage} 
\end{center}  
\caption{{\it The linear (quadratic)
mass-scale parameters $M, \tilde M$ are defined by $\delta c/c = E/M,
(E/{\tilde M})^2$,
respectively. The question marks in the Table indicate uncertain 
observational inputs. Hard limits are indicated by inequality signs.}}
\label{tabl1} 
\end{table}
}
\vspace{0.5cm}

Having established the existence of an observational opportunity, we now try to
give some flavour of the theoretical motivation for questioning the constancy
of the velocity of light, in the context of a quantum theory of gravity. A
preliminary remark is that any attempt to quantize gravity canonically must
involve a Lorentz-non-invariant separation of degrees of freedom and the choice
of a preferred reference frame. 
There is a natural local rest-frame in our
approximately Friedman-Robertson-Walker Universe, namely the comoving frame
identified approximately by the cosmic microwave background radiation. This
provides a natural frame in which to consider topological fluctuations in the
space-time background, as might arise from microscopic black holes or 
other
non-perturbative phenomena in quantum gravity -- 
the so-called space-time 
foam~(\cite{wheeler,hawking,hawking2,ehns,emn,garay,emnnew}).

Initially in the context of a string approach~(\cite{aemn}), it has been
argued that
foamy effects might lead the quantum-gravitational vacuum to
behave as a non-trivial medium, much like a plasma or other environment with
non-trivial optical properties. Another possible example 
of such behaviour has been proposed within a
canonical approach to quantum gravity~(\cite{pullin}, and it has also been
observed that
quantum fluctuations in the light-cone are to be expected~(\cite{ford2}).
The basic intuition
behind such suggestions is that quantum-gravitational fluctuations in the
vacuum must in general be modified by the passage of an energetic particle, and
that this recoil will be reflected in back-reaction effects on the propagating
particle itself.

Three possible optical effects of quantum gravity have been identified. One is
a simple energy-dependent reduction in photon velocity, namely a
frequency-dependent refractive index~(\cite{aemn,nature}). 
The second is a possible difference
between the velocities of photons of different helicities, namely
birefringence~(\cite{pullin}). An opportunity for an
experimental test of such a phenomenon has been provided recently by
the observations of polarized radiation from
GRB~990510~(\cite{polarC,polarW}).
The third is a possible energy-dependent diffusive spread in the
velocities of different photons of the same energy~(\cite{emnnew,ford,ford2}). 
In each case, the relevant
figure of merit would be (\ref{one}) if the effect is linear in
energy. 

Most of the rest of this paper constitutes a phenomenological analysis of
issues involved in the search for any such effect, and in disentangling it from
phenomena at the source, which we illustrate by a prototype analysis of the
available data on GRBs with established redshifts, with the 
results listed in Table 2 and 3.
Such an analysis has been made possible by the data from the
BeppoSAX satellite~(\cite{beppo}), 
which made it possible to observe afterglows from identified
GRBs, and measure their redshifts for the first time.

However, we first provide in Section 2 a simple analysis of the refractive
index induced by one particular recoil model for the quantum-gravitational
medium, which provides some physical flesh for the conceptual skeleton outlined
in this Introduction. In Sections 3 and 4 we discuss the propagation of
a pulse of electromagnetic radiation produced by an
ultra-relativistic source, as in a GRB explosion. Then, in
Section 5 we discuss how one may fit structures observed
in the available GRB data, and analyze their energy-dependence to look for
medium effects.
As we shall see, a key issue is to distinguish any such medium effects from
effects at the source. The former should increase when one considers GRBs at
larger distance $L$ or redshift $z$. We perform in Section 5 a prototype
regression analysis of the data from the handful of GRBs  whose redshifts are
currently known. Unsurprisingly, the present data do not exhibit any
significant correlation with $z$, but the method could usefully be
extended to
the hundreds  of GRBs whose redshifts are expected to be measured in coming
years. What if such a future analysis should yield a significant effect? It
should certainly not be believed without very critical review, and some of the
issues in the astrophysical modelling of GRBs and the correlation of data from
other sources such as AGNs are reviewed in the concluding Section 6.

\section{Quantum-Gravitational Recoil Effects on the Propagation of
Electromagnetic Waves} 

Our intuitive picture is that the quantum-gravitational
vacuum contains quantum fluctuations with typical sizes $\sim
l_P \sim 10^{-33}$~cm and time-scales $\sim t_P \sim
10^{-43}$~s. We then hypothesize that
particles propagating through the vacuum interact with these
fluctuations, inducing non-trivial recoil and
associated vacuum-polarization effects. As a specific
example, we have used the theoretical model of a
recoiling $D$ particle in the quantum-gravitational foam.
Its recoil due to scattering by a photon has been argued~(\cite{emndbrane})
to lead to a non-zero gravitational field 
with a metric of the form~(\cite{kanti}): 
\be
G_{ij} =\delta _{ij}, \; G_{00}=-1, \; G_{0i}=\epsilon^2(Y_i +
{\overline U}_i t)\Theta _\epsilon (t) 
\label{metrictarget}
\ee
where $0(i)$ denote time (space) components. 
Here, ${\overline U}_i$ is the recoil velocity of the
$D$ particle, which is located at $Y_i$, and
$\epsilon $ is a small parameter.
The metric (\ref{metrictarget}) implies that $D$-brane recoil 
induces the following perturbation $h_{\mu\nu}$ about flat space-time:
\be
h_{0i} = \epsilon ^2 {\overline U}_i t \Theta _\epsilon (t)
\label{pert2}
\ee
where we have only indicated 
the only non-zero components of $h_{\mu\nu}$.

We may identify $\epsilon^{-2} \sim t$
at large times, so that
the asymptotic form of the gravitational 
perturbation takes the form: 
\be
    h_{0i} \sim {\overline U}_i 
\label{asympt}
\ee
We note that this form corresponds to a breakdown of
Lorentz invariance induced by the propagation of the photon:
for symmetry reasons, any perturbation induced by
other models might be expected to take a similar form.
We suppose that the light travels
a distance $L$ in time $t \sim \epsilon^{-2}$ in  
the presence of a metric fluctuation $h_{0i}$ given by (\ref{asympt}).
The effects of such a field in Maxwell's equations
have been considered previously~(\cite{emnnew}): here we
present an elementary analysis, which does not require any
string or quantum field theory formalism. In fact,
is interesting to note that the interpretation
of Maxwell's equations
in the presence of a background gravitational field,
with a non-diagonal component $g_{0i}$ 
was proposed as an exercise for the reader by 
Landau and Lifshitz~(\cite{landau}), where a {\it formal} analogy with 
the propagation in a medium has been noted.

We parametrize the background metric in the form:
\be
    G_{00} \equiv -h, \qquad {\cal G}_i = - \frac{G_{0i}}{G_{00}}, \qquad i=1,2,3
\label{metricfour}
\ee
Maxwell's equations in this background metric in empty space 
can be written as~(\cite{landau}):
\bea 
&~& \nabla \cdot  B=0, \qquad \nabla \times H = \frac{1}{c} \frac{\partial }
{\partial t} D = 0, \nn \\
&~& \nabla \cdot D = 0, \qquad \nabla \times  E = -\frac{1}{c}\frac{\partial }
{\partial t} B = 0, 
\label{threemx}
\eea
where 
\be 
        D = \frac{E}{\sqrt{h}} + H \times {\cal G}, \qquad 
   B = \frac{H}{\sqrt{h}} + {\cal G} \times E 
\label{media}
\ee
Thus, there is a direct analogy 
with Maxwell's equations in a medium with $1/\sqrt{h}$ 
playing the r\^ole of the electric and magnetic permeability. 
In our case~(\cite{emndbrane}), $h=1$, so one has the 
same permeability as the classical vacuum. In the case of
the constant metric perturbation (\ref{asympt}),
after some elementary vector algebra and,
appropriate use of the modified Maxwell's equations,
the equations (\ref{threemx}) read: 
\bea
&~& \nabla \cdot E + {\overline U} \cdot {1 \over c}
\frac{\partial }{\partial t} E 
= 0        \nn \\
&~& \nabla \times B - \left(1 - {\overline U}^2\right) 
{1 \over c} \frac{\partial }{\partial t} E 
+ {\overline U} \times {1 \over c} \frac{\partial}{\partial t } B 
+ \left({\overline U} \cdot \nabla \right) E = 0 \nn \\ 
&~& \nabla \cdot B = 0 \nn \\
&~& \nabla \times E + \frac{1}{c}\frac{\partial}{\partial t}B = 0
\label{mxgrav}
\eea
Dropping non-leading terms of order ${\overline U}^2$
from these equations,
one obtains after some straightforward algebra
the following modified wave equations for $E$ and $B$: 
\bea
&~&\frac{1}{c^2}\frac{\partial ^2}{\partial ^2 t} B - 
\nabla ^2 B - 2 \left({\overline U}. \nabla \right) 
{1 \over c} \frac{\partial }{\partial t} B = 0 \nn \\
&~&\frac{1}{c^2}\frac{\partial ^2}{ \partial ^2 t} E - 
\nabla ^2 E - 2 \left({\overline U}. \nabla \right) 
{1 \over c} \frac{\partial}{\partial t} E = 0 
\label{waves}
\eea
If we consider one-dimensional motion along the $x$ direction,
we see that these equations admit wave solutions of the form
\be
E_x = E_z =0, \; E_y (x,t)= E_0 e^{ikx - \omega t}, \;\;
B_x = B_y =0, \; B_z (x,t) = B_0 e^{ikx - \omega t},
\label{wave}
\ee
with the modified dispersion relation:
\be
     k^2 - \omega ^2 - 2{\overline U} k \omega = 0
\label{dispr}
\ee
Since the sign of ${\overline U}$ is that of the momentum vector 
$k$ along the $x$ direction, the
dispersion relation (\ref{dispr}) corresponds to 
{\it subluminal} propagation with a refractive index: 
\be
c(E) =c \left(1 - {\overline U} \right)
+ {\cal O}\left({\overline U}^2\right)
\label{refr}
\ee
where we estimate that
\be
{\overline U} = {\cal O}\left(\frac{E}{M_Dc^2}\right)
\label{refrmag}
\ee
with $M_D$ the $D$-particle mass scale. This is
in turn given by $M_D = g_s^{-1} M_s$ in a string model,
where $g_s$ is the string coupling and $M_s$ is the string
scale~(\cite{emnnew}). 
The relation ({\ref{refrmag})
between ${\overline U}$ and the photon energy
has been shown~(\cite{lizzi,ms,emndbrane}) to
follow from a rigorous world-sheet analysis 
of modular divergences in string theory, but the details need not
concern us here. It merely expresses elementary
energy-momentum conservation.

The refractive index effect (\ref{refrmag}) is a mean-field effect,
which implies a delay in the arrival times of photons, relative
to that of an idealized low-energy photon for which
quantum-gravity effects can be ignored, of order:
\be
    \Delta t \sim \frac{L}{c} |{\overline U}| = 
{\cal O}\left(\frac{EL}{M_Dc^3}\right)
\label{refrmag2}
\ee 
We have discussed in some detail~(\cite{emnnew}) 
the quantum fluctuations about the mean-field
solution (\ref{refrmag2}), which would correspond 
in field theory to
quantum fluctuations in the light cone, and could be induced by
higher-genus
effects in a string approach. Such
effects would result in 
stochastic fluctuations in the velocity of light 
which are of order 
\be
\delta c \sim 8g_s E/M_Dc^2, 
\label{stochc}
\ee
where $g_s$ is the string coupling, which varies between ${\cal O}(1)$
and $ \ll 1$ in different string models. Such
an effect would
motivate the following parametrization of any possible
stochastic spread in photon arrival times:
\be
\left(\delta \Delta t \right) = {L E \over c\Lambda}
\label{stochc2}
\ee
where the string approach suggests that $\Lambda \sim M_Dc^2 / 8
g_s$.
We emphasize that,
in contrast to the variation (\ref{refrmag})
in the refractive index -
which refers to photons of different energy -
the fluctuation (\ref{stochc2}) 
characterizes the statistical spread in the velocities 
of photons {\it of the same energy}. 
Note that the stochastic effect (\ref{stochc2}) 
is suppressed, as compared to the refractive index 
mean field effect (\ref{refrmag}), by an extra power of $g_s$.

\section{Propagation of a Pulse of Photons through the
Space-Time Foam}

GRBs typically emit photons in pulses containing photons
with a combination of different wavelengths, whose sources
are believed to be ultrarelativistic shocks with Lorentz
factors $\gamma = {\cal O} (100)$~(\cite{Piran,Rees}). We do not
enter here into the details of the astrophysical modelling
of such sources. This is unnecessary for our present
exploratory study,
though it may be essential for future more detailed probes of
the constancy of the velocity of light. Instead, here we
study a simple generalization of
the previous discussion of monochromatic wave propagation,
considering a wave packet of photons emitted with a
Gaussian distribution in the light-cone variable $x - ct$.
Since the distance over which the ultrarelativistic source
moves during the emission is negligible compared with the
distance between the source and the observer, we may
represent the source equally well with a Gaussian distribution in $x$ at
the time $t=0$. This is adequate
to see how such a pulse would be modified at the observation point
at a subsequent time $t$, because of the propagation through the 
space-time foam, as a result of the refractive index effect 
(\ref{dispr}), (\ref{refr}). The phenomenon is similar to the motion 
of a wave packet in a conventional dispersive medium, as discussed 
extensively in the standard literature.

The Gaussian wavepacket may be expressed at $t = 0$ as the real part
of 
\be
f(x) = A e^{-x^2/(\Delta x_0)^2} e^{i k_0 x}
\label{gauss}
\ee
with a modulation envelope that is symmetrical about the origin, where it
has 
amplitude $A$. The quantity $\Delta x_0$ in (\ref{gauss}) denotes the 
root mean square of the spatial
spread of the energy distribution in the packet, which
is proportional to $|f(x)|^2$, as is well known.
If we assume a generic dispersion relation $\omega = \omega (k)$,
a standard analysis using Fourier transforms shows that 
at time $t$ the Gaussian wavepacket will have the form:
\be
   |f(x,t)|^2 = \frac{A^2}{(1 + \frac{\alpha ^2 t^2}{(\Delta x_0)^4})^{1/2}}  
e^{-\frac{(x - c_g t)^2 }{\left(2(\Delta x_0)^2
[1 + \frac{\alpha ^2 t^2}{(\Delta x_0)^4}]\right)}}
\label{timegauss}
\ee
where $\alpha \equiv \frac{1}{2} \left(d^2\omega /d^2 k\right)$,
and $c_g \equiv d\omega/dk$ is the group velocity.
This is the velocity with which the peak of the distribution moves
in time. 

We see immediately in (\ref{timegauss}) that the quadratic term 
$\alpha$ in the dispersion relation
does not affect the motion of the peak, but only the spread of the
Gaussian wave packet:
\be
 | \Delta x | = 
\Delta x_0 \left( 1 + \frac{\alpha ^2 t^2}{(\Delta x_0)^4}\right)^{1/2}
\label{spread}
\ee
which thus increases with time. The quadratic term $\alpha$ also affects 
the amplitude of the wave packet:
the latter decreases together with the increase in the spread
(\ref{spread}),
in such a way that the integral of $|f(x,t)|^2$ is constant. 

In the case of the quantum-gravitational foam scenario~(\cite{emnnew,aemn}), 
the dispersion relation 
assumes the following form for positive momentum $k$,
in units where $c=\hbar=1$: 
\bea
&~& k=\omega \left( 1 + \frac{\omega}{M_D}\right) 
~{\rm or} ~ \omega = k \left( 1 - \frac{k}{M_D} + \dots \right) \nn \\
&~&c_g = = \left(1 - {\overline U}\right) = 1 -
{\cal O}\left(\omega/M_D\right), \nn \\
&~&\qquad \alpha = - \frac{1}{M_D} + \dots 
\label{quant}
\eea
where we denote by $\dots$ denotes the higher-order (e.g., quadratic)
terms in $1/M_D$, which are subleading in this case.   
Thus the spread of the wave packet due to the non-trivial  
refractive index effect described in the previous section is:
\be
    | \Delta x| = \Delta x_0 
\left( 1 + \frac{t^2}{M_D^2 (\Delta x_0)^4} \right)^{1/2} 
\label{light}
\ee
We note that the spread due to the refractive index $\delta c/c \propto
\omega$ is independent of the energy of
the photon to leading order in $1/M_D$. 
We also note, therefore, that this effect is distinct from the
stochastic propagation effect, which gives rise to a spread 
(\ref{stochc2}) 
in the
wave-packet that depends on the photon energy $\omega$.
For astrophysical sources at cosmological distances
with redshifts $z \simeq 1$, and with an initial $\Delta x_0
$ of a few km, one finds that the correction (\ref{light})
is negligible if the quantum-gravity 
scale $M_D$ is of the order of $10^{19}$ GeV, namely of order 
$10^{-30}\Delta x_0$.
The correction would become of order $\Delta x_0$ only if the
latter is of order $10^{-3}$~m. Even if one allows $M_D$ to be as
low as the sensitivities shown in Table 1, this broadening effect
is still negligible for all the sources there, being at most of
order $10^{-22} \Delta x_0$. Therefore, in this particular
model, the only broadening effect that needs to be considered
is the stochastic quantum-gravitational effect on the 
refractive index that was introduced at the end of the previous section.

In the case of a quantum-gravitational foam scenario
with a quadratic refractive index: $\delta c / c \sim E^2$,
the dispersion relation assumes the following form:
\bea
&~& k=\omega \left( 1 + \frac{\omega}{\tilde{M}}\right) 
~{\rm or} ~ \omega = k \left( 1 - \left(\frac{k}{\tilde M}\right)^2 + 
\dots \right) \nn \\
&~&c_g = = \left(1 - {\overline U}\right) = 
1 - {\cal O}\left(\omega^2/{\tilde M}^2\right), \nn \\
&~&\qquad \alpha = - 3\frac{\omega}{{\tilde M}} + \dots 
\label{quant2}
\eea
where $\dots$ again denote subleading terms.   
In this case, 
the spread of the wave packet due to the non-trivial  
refractive index effect described above is:
\be
    | \Delta x| = \Delta x_0 
\left( 1 + \frac{9 \omega^2 t^2}{{\tilde M}^4 (\Delta x_0)^4} \right)^{1/2} 
\label{light2}
\ee
Once again, if one takes into account the sensitivities shown
in Table 1, the maximum spreading of the pulse is negligible
for $\Delta x_0 \sim 10^{-3}$~m, namely at most $\sim 10^{-33} \Delta
x_0$. Once again, one would need only to consider the possible
stochastic quantum-gravitational effect on the refractive
index. However, since a quadratic dependence is
not favoured theoretically, we do not pursue it further in the rest
of this paper.

\section{Cosmological Expansion and Light Propagation} 

We now discuss the implications of the cosmological expansion for the
searches for a quantum-gravity induced refractive index (\ref{refrmag2}) 
and a stochastic effect (\ref{stochc2}).
We work within the general context
of Friedman-Robertson-Walker (FRW) metrics, 
as appropriate for standard homogeneous and
isotropic cosmology~(\cite{wein}).
We denote by $R$ the FRW scale factor,
adding a subscript $0$ to denote the value at the present era,
$H_0$ is the present Hubble expansion parameter,
and the deceleration parameter
$q_0$ is defined 
in terms of the curvature $k$ of the FRW metric by
$k = (2q_0 - 1) (H_0^2 R_0^2 /c^2)$,
i.e., $\Omega _0 =2q_0$. 

Motivated by inflation and the cosmic microwave background data,
we assume a Universe with
a critical density: $\Omega _0=1$, $k=0$ and $q_0=1/2$.
We also assume that the Universe is matter-dominated
during all the epoch of interest.
Then the scale factor $R(t)$ of the Universe expands as:
\be
{R(t) \over R_0} = \left({3H_0 \over 2}\right)^{2/3} t^{2/3} 
\label{exp}
\ee
and the current age of the Universe is 
\be
 t_0 =\frac{2}{3 H_0}
\label{age}
\ee
Clearly no time delay can be larger than this.
The relation between redshift and scale factor is
\be
R(t)/R_0 =1/(1 + z) 
\label{redshift}
\ee
Substituting (\ref{redshift}) into (\ref{exp}),
we find the age of the Universe at any given redshift:
\be
   t=(\frac{2}{3H_0})\frac{1}{(1+ z)^{3/2}}=\frac{t_0}{(1+ z)^{3/2}}
\label{univage}
\ee
Hence, a photon (or other particle) emitted 
by an object at redshift $z$ has travelled for a time 
\be
 t_0 - t =\frac{2}{3H_0} \left(1 -\frac{1}{(1+z)^{3/2}}\right)
\label{tizero}
\ee
The corresponding differential relation between time and 
redshift is
\be
    dt = -\frac{1}{H_0}\frac{1}{(1 + z)^{5/2}}dz 
\label{diff}
\ee
This means that 
during the corresponding infinitesimal time (redshift) interval,
a particle with velocity $u$ travels a distance
\be
u\,dt = -\frac{1}{H_0}\frac{u}{(1 + z)^{5/2}}dz.
\ee
Therefore, the total distance $L$ travelled by such a
particle since emission at redshift $z$ is
\be 
     L = \int_t^{t_0} u dt = 
\frac{1}{H_0}\int _0^z \frac{u(z)}{(1 + z)^{5/2}}dz 
\label{distance}
\ee
hence, the difference in distances covered by two particles
with velocities differing by $\Delta u$ is: 
\be
   \Delta L =\frac{1}{H_0} \int _0^z \frac{dz}{(1+ z)^{5/2}}(\Delta u)
\label{du}
\ee
where we allow $\Delta u$ to depend on $z$.

In the context of our quantum-gravity-induced refractive-index phenomenon
(\ref{refrmag}), we are confronted with just such a situation.
Consider in that context two photons travelling with
velocities very close to $c$,
whose present-day energies are $E_1$ and $E_2$. 
At earlier epochs, their energies would have been blueshifted
by a common factor $1 + z$. Defining $\Delta E_0 \equiv E_1 - E_2$,
we infer from (\ref{refrmag}) that $\Delta u = (\Delta E_0 \cdot (1 +
z))/M$.
Inserting this into (\ref{du}), 
we find an induced difference
in the arrival time of the two photons given by
\be
  \Delta t =\frac{\Delta L}{c} \simeq \frac{2}{H_0}\left[ 1 - \frac{1}{(1 + z)^{1/2}}\right] 
\frac{\Delta E_0}{M}. 
\label{deltacosm}
\ee
The expression (\ref{deltacosm}) describes the corrections 
to the refractive index effect (\ref{refrmag2})
due to the cosmological expansion. 
For small $z << 1$, the general expression (\ref{deltacosm}) 
yields $\Delta t \simeq (z \cdot \Delta E_0)/ (H_0 \cdot M)$,
which agrees with the simple expectation
$\Delta t \simeq (r \cdot \Delta E_0) / (c \cdot M)$
for a nearby source at distance $r = c(t_0 - t) \simeq
{z}/{H_0} + \dots $.  
There would be similar cosmological corrections to the 
stochastic effect (\ref{stochc2}), also
given by an expression of the form (\ref{deltacosm}),
but with $\Delta E_0 \rightarrow E$, $M \rightarrow \Lambda$,
where $E$ is a typical energy scale in a single channel.

In the next Section we present a detailed 
analysis of the astrophysical data for the five GRBs 
listed in Section 2, whose redshifts $z$ are known.
We shall be looking for a correlation with the redshift,
calculating a regression measure for the 
effect (\ref{deltacosm}) and
its stochastic counterpart. 
Specifically, we shall concentrate on
looking for linear dependences of
the ``observed'' $\Delta t/\Delta E_0$ and the spread 
$\Delta \sigma/E$ on ${\tilde z}
\equiv 2 \cdot [ 1 - (1 / (1 + z)^{1/2}] \simeq z - (3/4) z^2 +
\dots$.

\section{GRB Data Analysis} 

We present in this section a model analysis of
astrophysical data on GRB pulses that allows us to place a bound on
the phenomena discussed in the previous sections, motivated by
possible non-trivial medium effects of quantum gravity on the 
propagation of photon probes. Previously, the data from
individual GRBs~(\cite{nature,schaefer}), AGNs~(\cite{biller}) and
pulsars~(\cite{crab}) had been considered.
Here we take a further step, analyzing the data from those
GRBs whose redshifts are known after identification of their
optical counterparts. This enables us to perform a regression
analysis to seasrch for a possible correlation with redshift (distance),
as a first attempt to unravel source and medium effects.
However, this analysis
should only be considered as a prototype. Much more sophisticated fits
to the data of individual GRBs could be attempted, but this would
probably be worthwhile only when many more GRB redshifts are known.
We anticipate that this should be the case within a year or two.
It should also be emphasized that, even if an effect correlated
with redshift were to be detected, it would still be necessary
to confirm that it was a medium effect rather than an evolutionary
effect in the GRB sources. This would surely require detailed
astrophysical sources and/or a confirmation of a similar effect
in emissions from a different population of distant
astrophysical sources, such as AGNs. However, these topics lie
beyond the scope of the present article. 

In our searches for the effects
described in Sections 2 and 3, we look 
for short-duration structures in the time profiles 
of those GRBs whose redshifts, and hence distances, are known 
with some precision. We then make
appropriate fits of the astrophysical data 
in various energy channels, seeking to constrain differences
in the timings and widths of peaks for different energy
ranges. Simultaneity of the 
peak arrival times at different energies would place bounds on the
induced refractive index (\ref{refr}) of photons. Independence
of the widths of peaks from the channel energies would
constrain stochastic fluctuations in the velocities of
photons of the same energy.

The sample of GRB data discussed 
below have been taken from the BATSE catalogue~(\cite{batse})
and OSSE data~(\cite{osse}).
We focus on the following five GRBs, whose redshifts $z$
are known:  

\begin{itemize} 

\item GRB~970508 with BATSE trigger number 6225 and redshift $z$=0.835,

\item GRB~971214  with BATSE trigger number 6533 and redshift $z$=3.14,

\item GRB~980329 with BATSE trigger number 6665 and redshift $z$=5.0,

\item GRB~980703  with BATSE trigger number 6891 and redshift $z$=0.966,

\item GRB~990123  with BATSE trigger number 7343 and redshift $z$=1.60.

\end{itemize}

\noindent
We recall the energy ranges in which BATSE generally observes
photons: Channel~1 in the energy range (20,50) keV,
Channel~2  between (50,100) keV, Channel~3 between (100,300) keV and 
Channel~4 above 300~keV. We note that the energies recorded by
BATSE are not the exact photon energies, and that there is
in particular some feedthrough from high-energy photons into
lower-energy channels. This effect can be neglected in the
pioneering analysis that we undertake here, 
and is sidestepped in our later comparison of BATSE and OSSE data, but may
need to
be taken into account in any more detailed follow-up analysis.

The data for each GRB exhibit non-trivial and non-universal
structures in time. For each of the triggers studied, we
have fitted one or two of the prominent peaks in each of the energy
channels, with the aim of looking for or constraining their
differences in time or width between different energy channels.
We have explored four different functions for the fits to the different
peaks:
(i) a Gaussian function characterized by the peak location $t_p$
and width parameter $\sigma$, (ii) a
Lorentzian function characterized by:
$A/[(t-t_p)^2 + (\Gamma/2)^2] $
(iii) a `tail' fit with fitting function 
\be
  N(t)=c_1*(t-t_0)^m~exp[-(t-t_0)^2/(2\cdot\tau^2)], \quad t>t_0 
\label{fit2}
\ee 
\nk
which peaks at $t_p=\tau\sqrt{m}+t_0$,
to take into account the tail that tends to appear in the data 
after the peak,
and (iv) the phenomenological `pulse' model of Norris et al.,
which has the functional form~(\cite{norris})  
\be
 N(t)=c_1~{\rm exp}\left[-({\rm abs}(t-t_p)/\sigma_{r,d})^\nu\right]
\label{fit3}
\ee
where $t_{p}$ is the time at which the photon pulse takes its maximum,
$\sigma_r$ and $\sigma_d $ 
are the rise and decay times of the distribution,
respectively, and $\nu$ gives the sharpness or smoothness of the pulse
at its peak.

We compare in Fig.~1
the four fits to the data for GRB 970508 in Channels 1 and 3.
As seen in Fig. 1, the Gaussian and Lorentzian fits are of lower
quality than the `tail' and `pulse' fits,
so we
concentrate on the latter for the remaining GRBs.
Figs.~2 to 5 show the `tail' and `pulse' fits for the remaining GRBs
that we study: 971214, 980329, 980703 and 990123, respectively.
We compile in Table 2 the values of the `tail' and `pulse' fit parameters 
that we find for all the GRBs. Specifically, we list 
for both the `tail' and `pulse' fitting functions the peak
time, $t_p$, and the pulse width, $\sigma$, defined as 
half of the width of the pulse at $e^{-1/2}\simeq 60\%$ of its maximum 
value. When applied to the `tail' distribution
(\ref{fit2}), this definition yields 
\be
\sigma = \alpha \tau \sqrt{m},
\label{sigmatail}
\ee
where $\tau$ and $m$ are defined in (\ref{fit2}), and
$\alpha > 0$ is the solution of the equation    
\be
       {\rm ln}(1 + \alpha)-\frac{1}{2}(1 + \alpha )^2 + \frac{1}{2}(1
       + m^{-1}) =0 ,
\label{alphaeq}
\ee
whilst for the `pulse' distribution 
this definition yields $\sigma = (\sigma_r + \sigma_d)/2$.

The values recorded in Table \ref{table2} may be of more
general interest to those modelling GRBs. However, our main
interest here is to compare the values
of these parameters in the different channels, and use their
differences to constrain energy-dependent differences and
stochastic fluctuations in photon velocities.
As seen in Table \ref{table2}, 
we find that the different fitting functions yield constraints 
on the propagation parameters 
that are
comparable within the statistical errors: we use the differences
between them as gauges of the systematic errors. 

The only candidate that we see for a systematic trend in the data
is a tendency for pulses in the higher-energy channels to be
narrower than in the lower-energy channels.
This effect is seen clearly in Fig.~1 for the case of GRB 970508.
However, this narrowing is the opposite of what we would suggest
theoretically, which would be a slowing and broadening of the
peak at higher energies.

We also compare data from Channel 3  of the BATSE detector 
with the data from OSSE detector~(\cite{osse}), which
detects photons in a single channel
with energy range $1 < E < 5-10$ MeV. Since the OSSE
data are at higher energies, they are more sensitive to
the type of energy-dependent effect of interest to us.
The reason we compare the OSSE data with Channel 3 
of the BATSE data is that the latter are free of contamination
by the data in lower-energy channels, removing one particular
possible source of systematic error.
OSSE data are available for the GRBs 980329 and 990123, which
we display in Figs.~\ref{fig6} and \ref{fig7}, respectively. 
The results of our numerical analysis of the arrival times
and widths of identified
OSSE pulses are given in Table~\ref{table3}.

In order to investigate the possible 
fundamental physics significance of this or any other possible
energy-dependent effect, we have compiled the data from all the
GRBs we have studied as functions of the figure of merit
${\tilde z}=1 - 1/(1 + z)^{1/2}$ introduced at the end of
Section 4, as seen
in Fig.~8 for the locations of the peaks, and in Fig.~9 for the
width parameters. 
In some cases, there are two BATSE points with the
same redshift, reflecting the fact that we have fitted two peaks in the
BATSE data for the corresponding GRB. Also plotted are the results
of the OSSE analysis for the GRBs 980329 and 990123.
The inner error bars in Figs.~8 and 9
are the purely statistical errors produced by the fitting routines,
taking as central values those extracted from the `pulse' fits, which
we consider to be the most reliable.
The outer error bars are obtained by adding in quadrature a
theoretical `systematic' error, defined by the differences between
the values of the fitting parameters obtained from the `tail' and `pulse'
fits. All the numbers used are taken from Tables 2 and 3.

No strong correlation with $z$ is apparent in either Fig.~8 or Fig.~9.
We have performed a regression analysis for linear dependences of
the forms
\begin{equation}
y({\tilde z})=a {\tilde z} + b
\label{lineardep}
\end{equation}
for both the time delays $y = t_p$ and the width parameters
$y = \sigma$.

The extracted regression coefficients,
defined by
$r^2 \equiv (\sum _{i} \left( y_{\rm est} - {\overline y}
\right)^2) / (\sum _{i} 
\left( y - {\overline y} \right)^2)$,
where $y_{\rm est}$ is the estimated value given by (\ref{lineardep}) 
and ${\overline y}$ is the mean value of the experimental data, are
\be
{\rm Time~Delays}: \;\; r^2_{t_p} = 0.11, \qquad
{\rm Widths}: \;\; r^2_\sigma = 0.12,
\label{regressionnumbers}
\ee
indicating no significant correlation. As an exercise, we
have repeated the regression analysis omitting individual
data points, to see whether any rogue point could be concealing
a significant effect: again, no significant correlation was found.

The formal results of the linear fits
shown in Figs.~\ref{figregz} and \ref{figregwidth}
for the coefficients defined in (\ref{regressionnumbers}) are
\bea
{\rm Time~Delays}: \quad &~&a=-3.0(5),~~ b=1.4(2) 
 , \nn \\
{\rm Widths}: \quad &~&a=-1.6(6),~~ b=0.3(2).
\label{fitresults}
\eea
The negative slopes of these fits are opposite in sign to those
expected for the quantum-gravity refractive-index
effect (\ref{deltacosm}) and its stochastic fluctuations.
As already discussed, the regression analysis indicates that
neither of the $a$ values should be interpreted as a real
effect, because of the scatters in the data sets.
We determine limits on
the quantum gravity scales $M$ and $\Lambda$ by
identifying the magnitudes of the slopes $a$ in (\ref{fitresults}) with
the coefficients of the ${\tilde z}$ terms in (\ref{deltacosm})
and its equivalent for the width parameter.
Using the current value for the Hubble expansion parameter,
$H_0 = 100 \cdot h_0$~km/s/Mpc, where $0.6 < h_0 < 0.8$,
we obtain from the the regression fits of Figs.~\ref{figregz} and 
\ref{figregwidth} the following limits
\be 
M \gsim 10 ^{15} \; {\rm GeV}, \quad \Lambda \gsim 2 \times 10 ^{15} \;
{\rm GeV} 
\label{limit}
\ee
on the possible quantum-gravity effects.

\begin{figure}[htb]
\epsfxsize=\linewidth 
\bigskip
\centerline{\epsffile{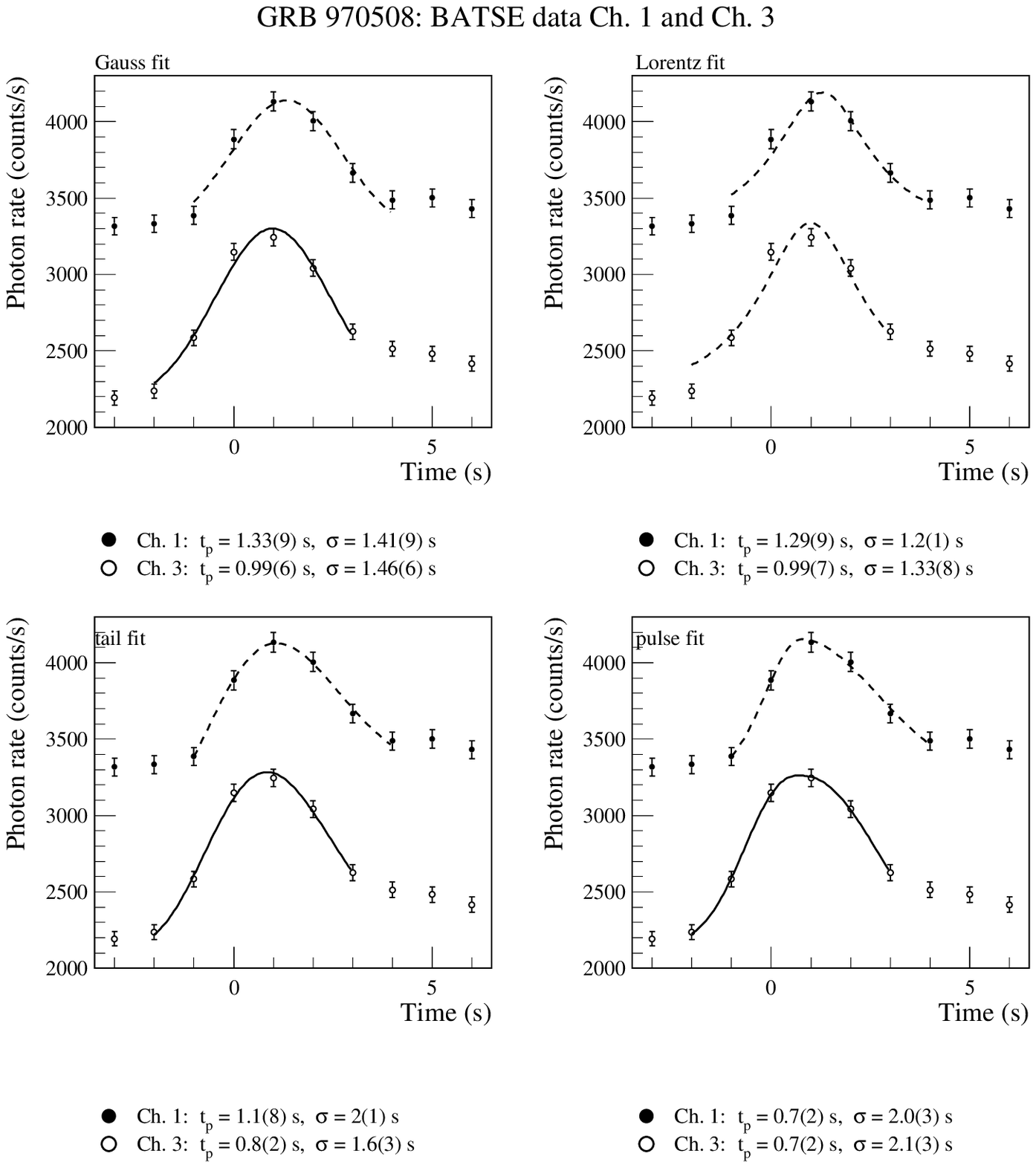}}
\caption{\it\baselineskip=12pt Time distribution 
of the number of photons 
observed by BATSE in Channels 1 and 3 for GRB~970508, compared 
with the following fitting functions: (a) Gaussian, (b)
Lorentzian, (c) `tail' function, and (d) `pulse' function.
We list below each panel the positions $t_p$ and widths $\sigma_p$
(with statistical errors) found for each peak in each fit. We
recall that the BATSE data are binned in periods of 1.024~s.}

\bigskip
\label{fig1}\end{figure}

\begin{figure}[htb]
\epsfxsize=\linewidth
\bigskip
\centerline{\epsffile{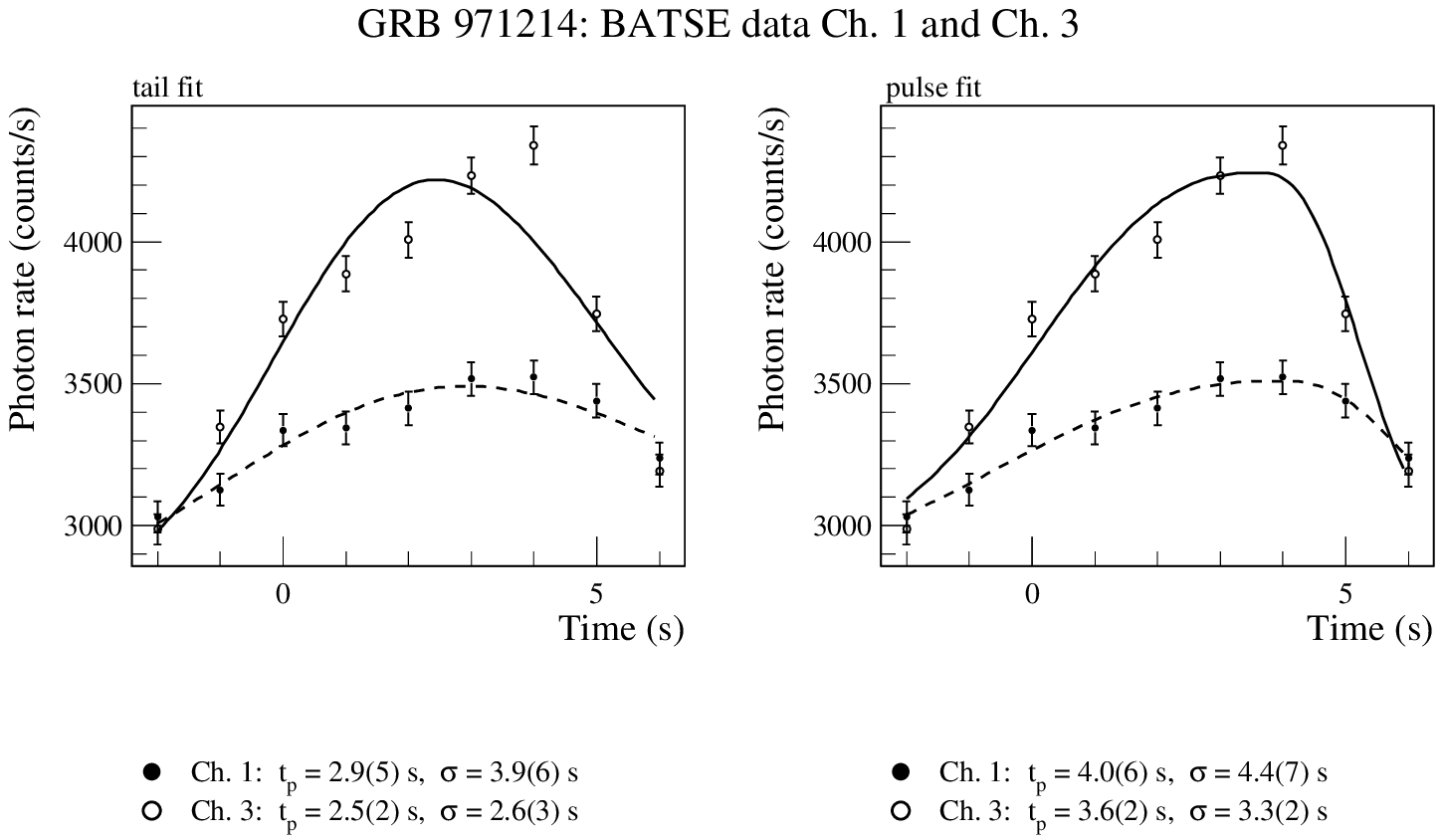}}
\caption{\it\baselineskip=12pt 
Time distribution of the number of photons
observed by BATSE in Channels 1 and 3 for GRB~971214, compared
with the following fitting functions: (a) `tail' function, and (b) `pulse'
function.
We list below each panel the positions $t_p$ and widths $\sigma_p$
(with statistical errors) found for each peak in each fit.}
\bigskip
\label{fig2}\end{figure}

\begin{figure}[htb]
\epsfxsize=\linewidth
\bigskip
\centerline{\epsffile{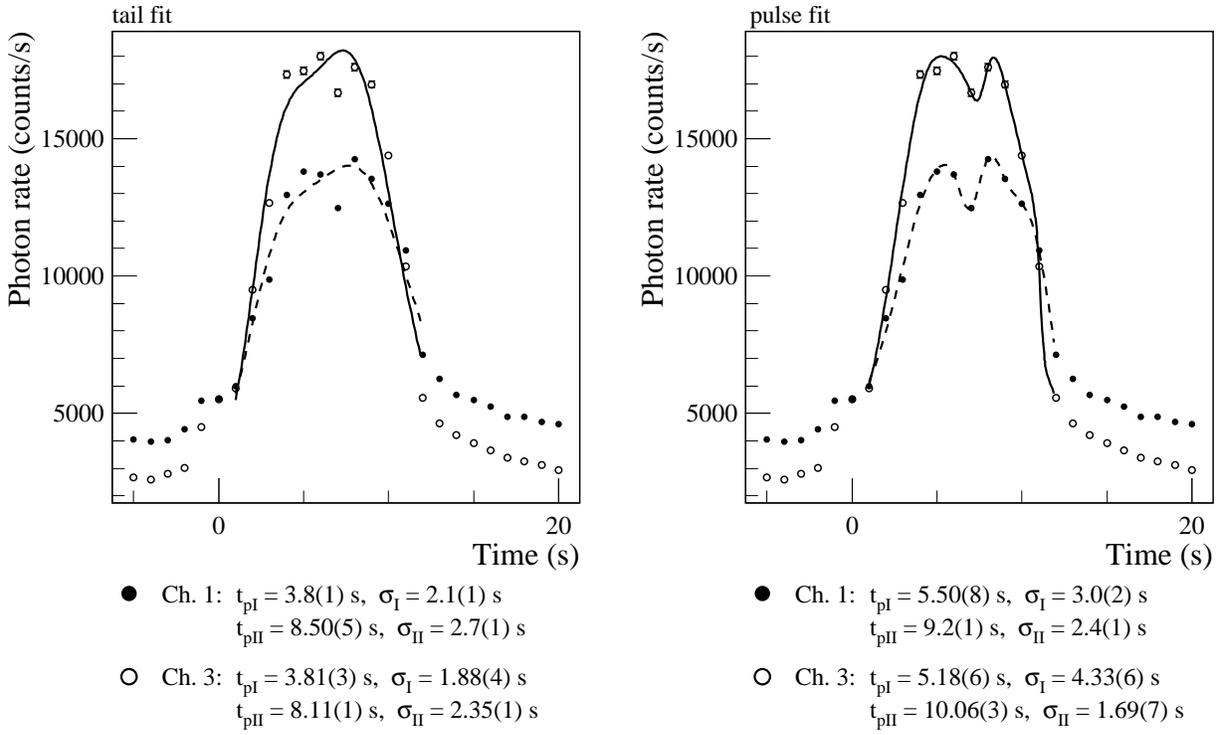}}
\caption{\it\baselineskip=12pt 
As in fig. \ref{fig2}, but for GRB~980329.} 
\bigskip
\label{fig3}\end{figure}

\begin{figure}[htb]
\epsfxsize=\linewidth
\bigskip
\centerline{\epsffile{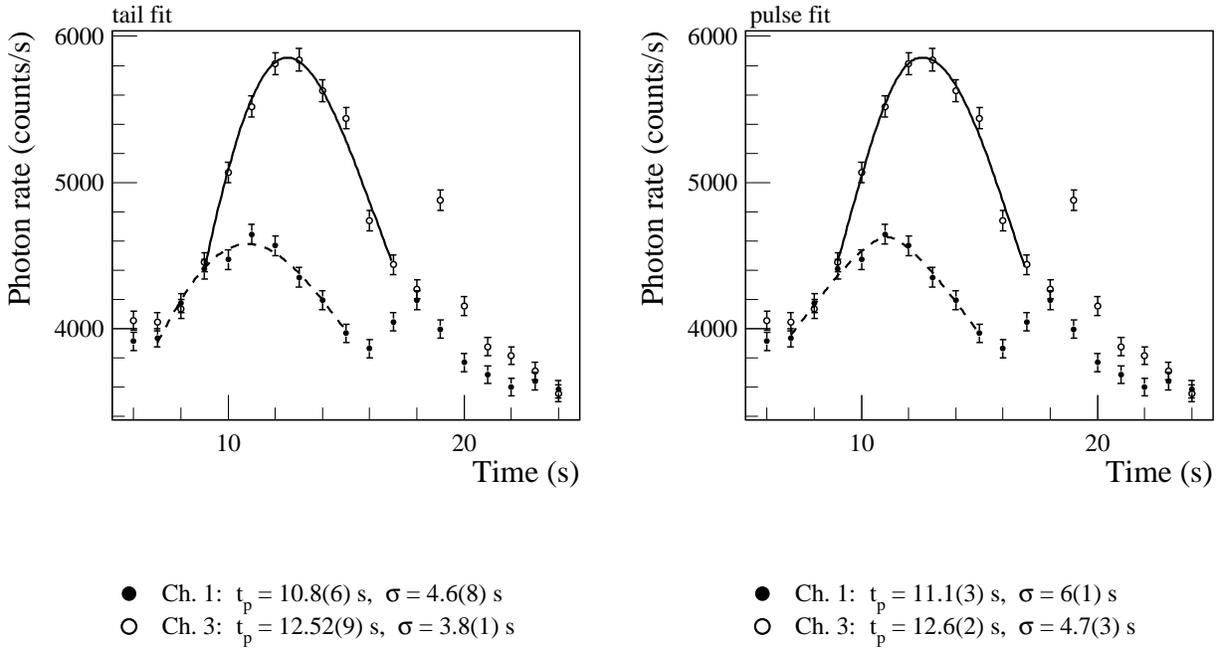}}
\caption{\it\baselineskip=12pt
As in fig. \ref{fig2}, but for GRB~980703.}  
\bigskip
\label{fig4}\end{figure}

\begin{figure}[htb]
\epsfxsize=\linewidth
\bigskip
\centerline{\epsffile{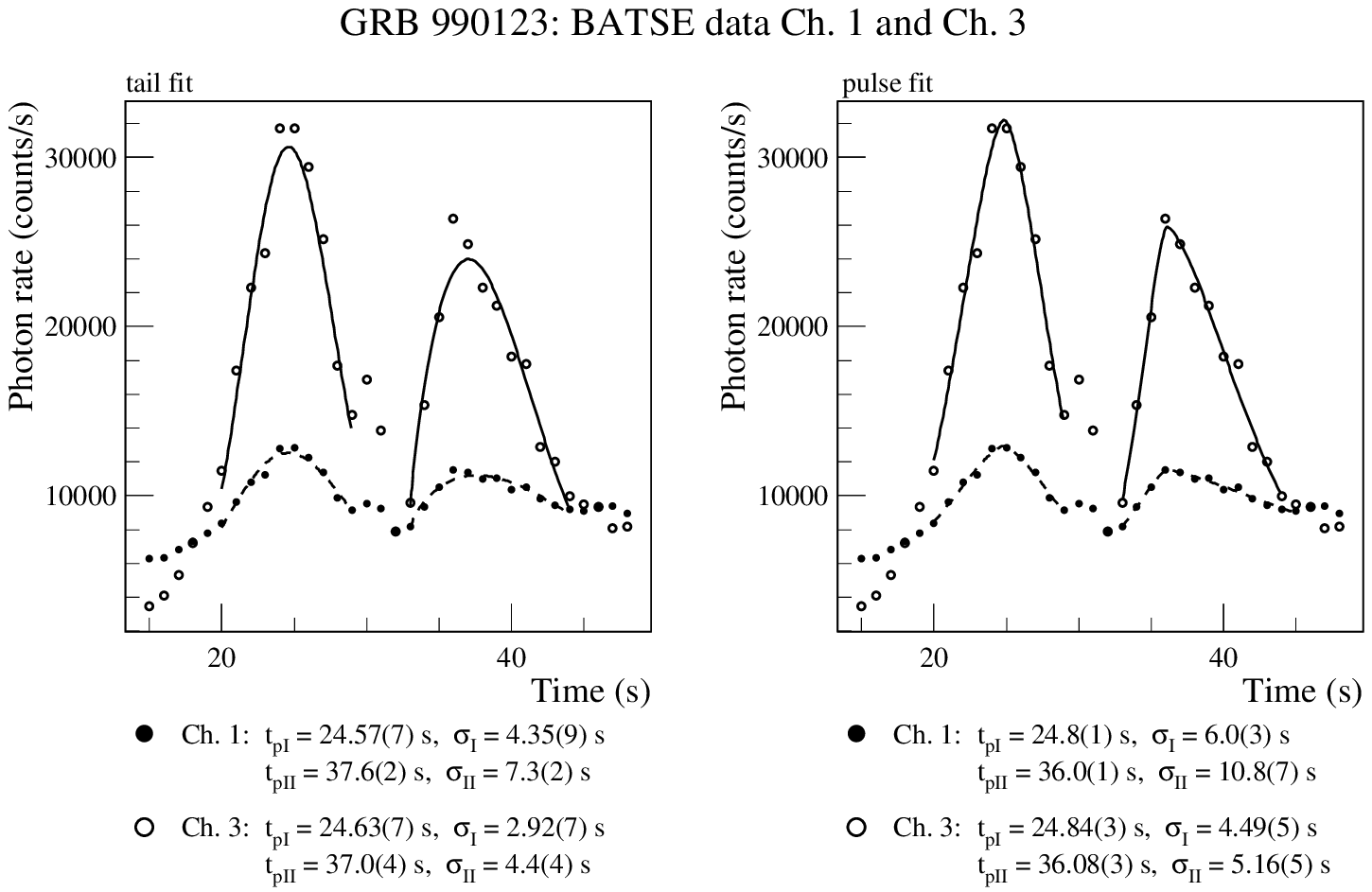}}
\caption{\it\baselineskip=12pt 
As in fig. \ref{fig2},
but for GRB~990123.}  
\bigskip
\label{fig5}\end{figure}

\begin{figure}[htb]
\epsfxsize=\linewidth
\bigskip
\centerline{\epsffile{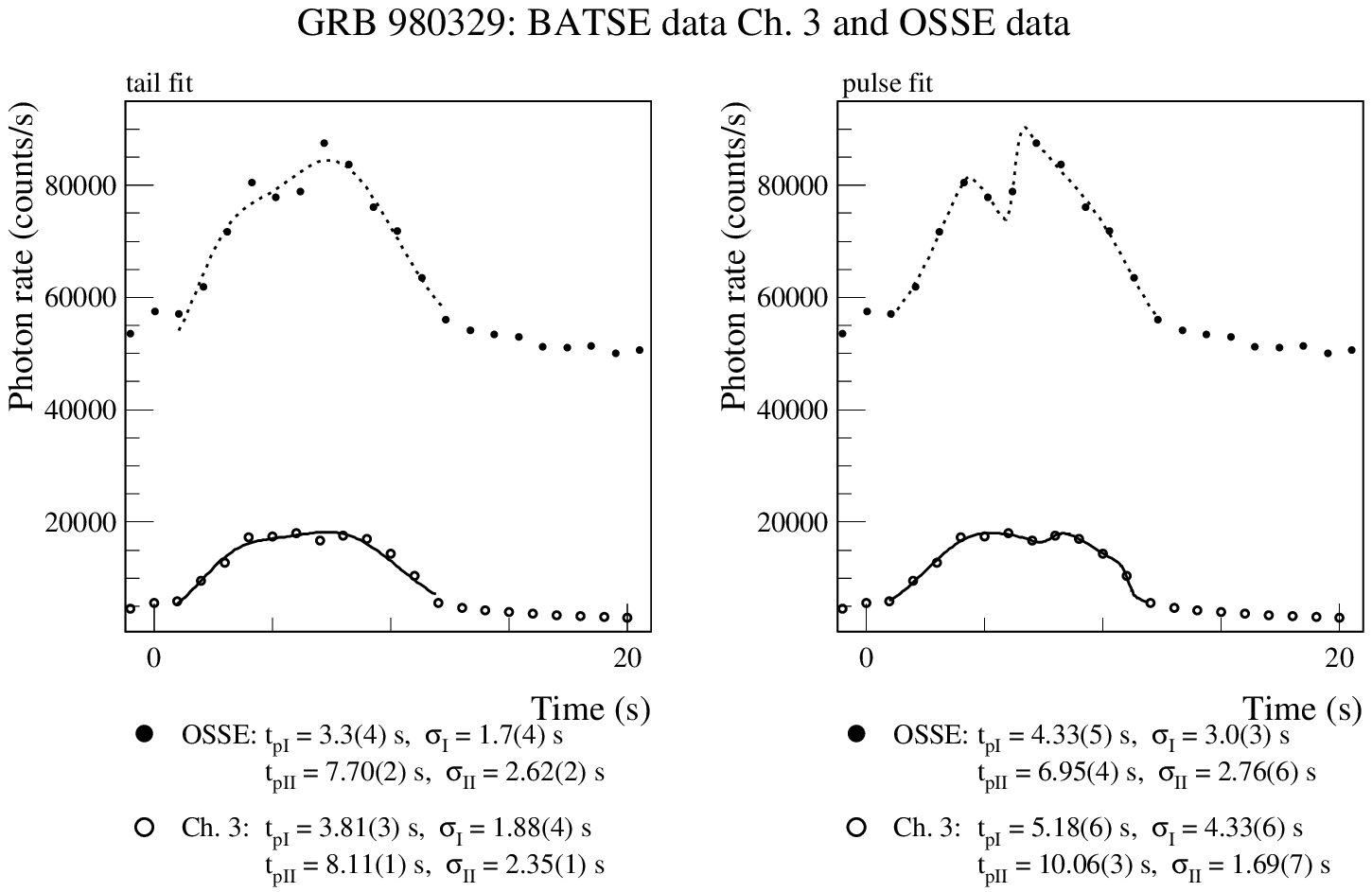}}
\caption{\it\baselineskip=12pt 
Time distribution of the number of photons
observed by OSSE and by BATSE in Channel 3 for GRB~980329, compared
with the following fitting functions: (a) `tail' function, and (b) `pulse'
function.
We list below each panel the positions $t_p$ and widths $\sigma_p$
(with statistical errors) found for each peak in each fit.}
\bigskip
\label{fig6}\end{figure}

\begin{figure}[htb]
\epsfxsize=\linewidth
\bigskip
\centerline{\epsffile{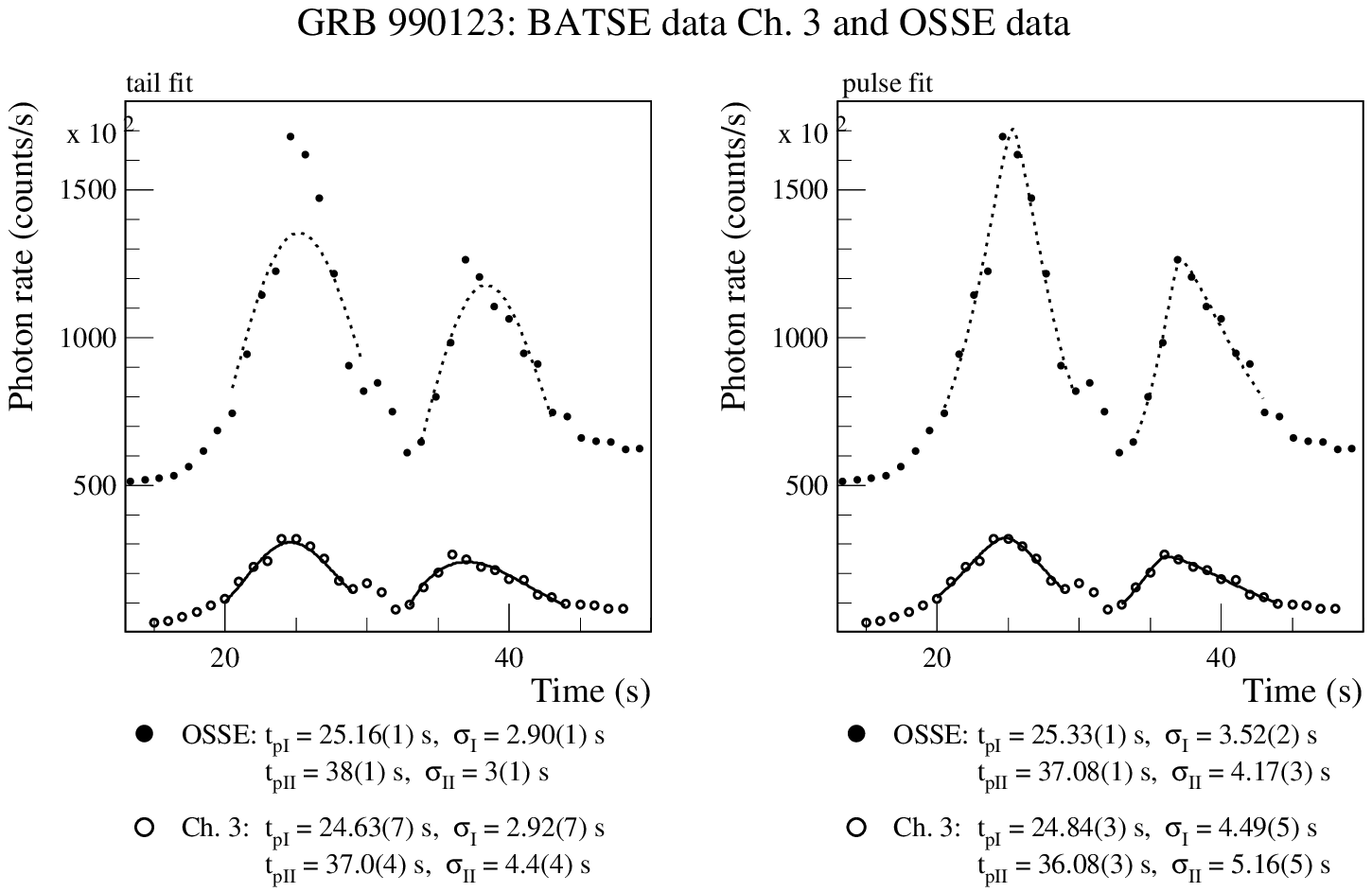}}
\caption{\it\baselineskip=12pt 
As in Fig.~\ref{fig6}, but for GRB~990123.}
\bigskip
\label{fig7}\end{figure}

\begin{figure}[htb]
\epsfxsize=\linewidth
\bigskip
\centerline{\epsffile{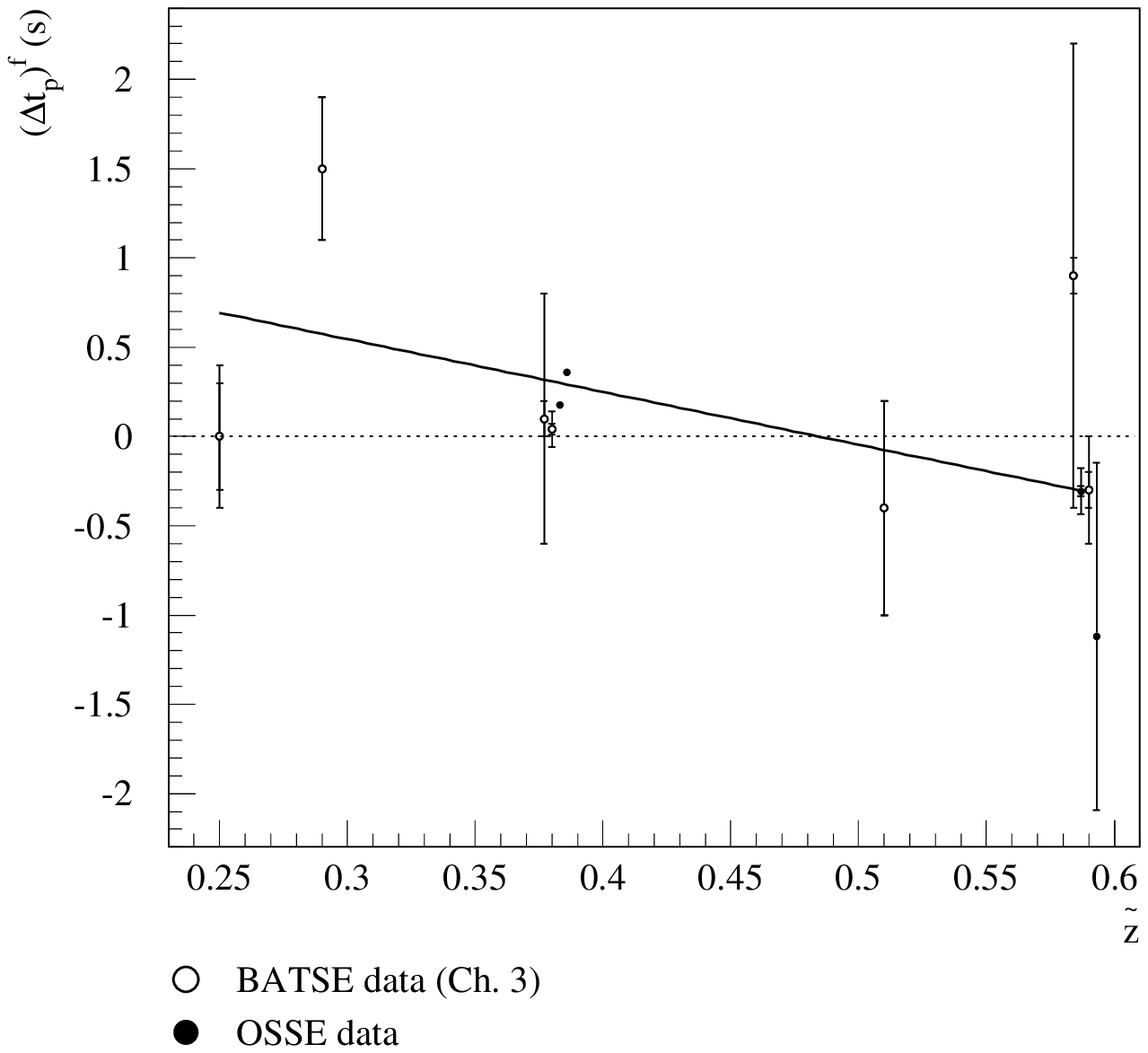}}
\caption{\it\baselineskip=12pt Values of the shifts $(\Delta t_p)^f$ in
the timings of the peaks fitted for each GRB studied
using BATSE and OSSE data, plotted versus 
${\tilde z}=1-(1+z)^{-1/2}$, where $z$ is the 
redshift. The indicated errors are the
statistical errors in the `pulse' fits provided by the fitting routine,
combined with systematic error estimates obtained
by comparing the results obtained using the `tail' fitting
function. The values obtained by comparing OSSE with BATSE 
Channel 3 data
have been rescaled by the factor $(E_{min}^{BATSE~Ch.~3} -
E_{max}^{BATSE~Ch.~1}) / (E_{min}^{OSSE} - E_{max}^{BATSE~Ch.~3})$,
so as to make them directly comparable with the comparisons of
BATSE Channels 1 and 3. The solid line is the best linear fit.}
\bigskip
\label{figregz}\end{figure}

\begin{figure}[htb]
\epsfxsize=\linewidth 
\bigskip
\centerline{\epsffile{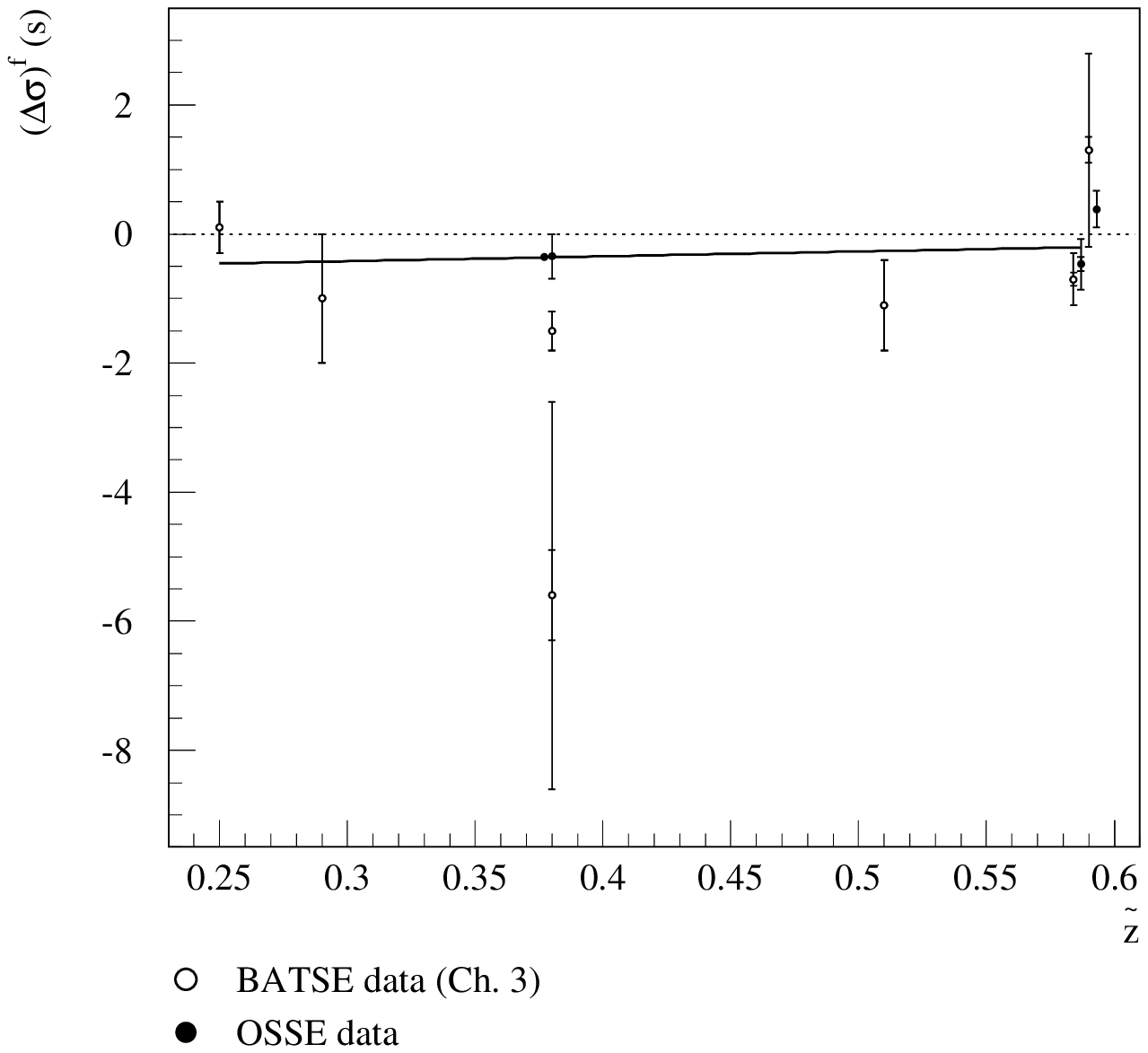}}
\caption{\it\baselineskip=12pt 
Values of the changes $(\Delta \sigma)^f$ in
the widths of the peaks fitted for each GRB studied
using BATSE and OSSE data, plotted versus
${\tilde z}=1-(1+z)^{-1/2}$, where $z$ is the
redshift. The indicated errors are the
statistical errors in the `pulse' fits provided by the fitting
routine,
combined with systematic error estimates obtained
by comparing the results obtained using the `tail' fitting
function. The values obtained by comparing OSSE with BATSE
Channel 3 data
have been rescaled by the factor $(E_{min}^{BATSE~Ch.~3} -
E_{max}^{BATSE~Ch.~1}) / (E_{min}^{OSSE} - E_{max}^{BATSE~Ch.~3})$,
so as to make them directly comparable with the comparisons of
BATSE Channels 1 and 3. The solid line is the best linear fit.}
\bigskip
\label{figregwidth}\end{figure}

{\scriptsize
\begin{table}[ht] 
\begin{center}
\begin{tabular}{|c|c|c||c|c|c|c|c|c|c|}   \hline
& & & \parbox{1cm}{GRB 970508} & \parbox{1cm}{GRB 971214} & 
\parbox{1cm}{GRB 980329 (I)}  & \parbox{1cm}{GRB 980329 (II)} & 
\parbox{1cm}{GRB 980703} & \parbox{1cm}{GRB 990123 (I)} & 
\parbox{1cm}{GRB 990123 (II)} 
\\ \hline \hline 
Ch 1& 
Tail & $t_p$ (s) & 1.1(8) & 2.9(5) & 3.8(1) & 8.50(5) & 10.8(6) & 24.57(7)
   & 37.6(2)  \\ \hline
& & $\sigma$ (s) &2(1)  &3.9(6)  &2.1(1)  &2.7(1)  &4.6(8)  &4.35(9) &7.3(2)   
\\ \hline\hline 
& & $t_p$ (s) & 0.7(2) & 4.0(6) & 5.50(8) & 9.2(1) & 11.1(3) & 24.8(1)
   & 36.0(1)    \\ \hline
& Pulse & $\sigma_r$ (s) & 1.3(3) & 5.8(3) & 3.65(9) & 1.9(1) &7(1) & 6.1(2)
   & 4.6(2)   \\ \hline
& & $\sigma_d$ (s) & 2.7(3) & 3(1) & 2.3(2) & 2.8(1) & 6(1) & 6.0(3)
   & 17(1)  \\ \hline
& & $\sigma$ (s) & 2.0(3) & 4.4(7) & 3.0(2) & 2.4(1) & 6(1) & 6.0(3) 
   & 10.8(7)  \\ \hline \hline 
Ch 3& Tail & $t_p$ (s) & 0.8(2) & 2.5(2) & 3.81(3) & 8.11(1) & 12.52(9) & 24.63(7)
   & 37.0(4) \\ \hline
& & $\sigma$ (s) &1.6(3)  &2.6(3)  &1.88(4)   &2.35(1)  &3.8(1)  
&2.92(7)  &4.4(4)
\\ \hline\hline
& & $t_p$ (s) & 0.7(2) & 3.6(2) & 5.18(6) & 10.06(3) & 12.6(2) & 24.84(3)
   & 36.08(3)    \\ \hline
& Pulse & $\sigma_r$ (s) & 1.8(3) & 4.6(2) & 3.37(6) & 2.3(1) & 4.3(3) 
   & 4.59(4) & 2.85(4)   \\ \hline
& & $\sigma_d$ (s) & 2.5(3) & 2.1(2) & 5.28(6) & 1.071(7) & 5.2(3) & 4.39(5) 
   & 7.48(5)   \\ \hline
& & $\sigma$ (s) & 2.1(3) & 3.3(2) & 4.33(6) & 1.69(7) & 4.7(3) & 4.49(5) 
   & 5.16(5)  \\ \hline \hline 
$\Delta$ & Tail & $\Delta t_p$ (s) & -0.3(8) & -0.4(5) & 0.0(1) & -0.39(5) 
   & 1.7(6) & 0.1(1) & -0.6(4)   \\ \hline
& & $\Delta\sigma$ (s)  &0(1)  &-1.3(7)  &-0.2(1)  &-0.3(1)  &-0.8(8)  
&-1.4(1)  & -2.9(4)  
\\ \hline \hline
& Pulse & $\Delta t_p$ (s) & 0.0(3) & -0.4(6) & -0.3(1) & 0.9(1) & 1.5(4) 
  & 0.04(3) & 0.1(1)    \\ \hline
& & $\Delta\sigma$ (s) &0.1(4)  &-1.1(7)  & 1.3(2)  &-0.7(1)
 &-1(1)  &-1.5(3) & -5.6(7)  \\ \hline \hline
& & $(\Delta t_p)^{{\rm f}}$ (s) &0.0(3)(3) &-0.4(6)(0) &
-0.3(1)(3) &0.9(1)(13) &1.5(4)(2) &0.04(3)(10) & 0.1(1)(7)\\ \hline \hline
& & $(\Delta\sigma )^{\rm f}$ (s) &0.1(4)(1)  &-1.1(7)(2)  &
1.3(2)(15)  &-0.7(1)(4) &-1(1)(0.2) &-1.5(3)(1) &-5.6(7)(27)   \\ \hline 
\end{tabular} 
\end{center} 
\caption{{\it Results of fits to the GRB data from BATSE.
For $A=t_p, \sigma$ in each case,  
the quantities 
$\Delta A$ denote the differences
$\Delta A \equiv A^{{\rm Ch.3}}- A^{{\rm Ch.1}}$, and
the quantities $(\Delta A)^{{\rm f}} \equiv 
(\Delta A)^{pulse} ( \sqrt{\delta A_{(1)}^2+ \delta A_{(3)}^2} ) 
( \delta(\Delta A))$ are our
final results for the induced 
time delays or width differences
of the peaks between BATSE Channels 1 and 3.
The first parenthesis 
in the latter expression denotes the statistical error (where
$\delta A_{(i)}$ denotes the statistical error in determining $A$ in the
$i^{th}$ channel), whilst the second parenthesis
denotes the theoretical ``systematic'' error, defined as
$\delta (\Delta A) \equiv 
\left|(\Delta A)^{pulse}  -(\Delta A)^{tail}\right|$. 
}}
\label{table2} 
\end{table}
}

{\scriptsize
\begin{table}[ht] 
\begin{center}
\begin{tabular}{|c|c|c||c|c|c|c|}   \hline
& & & 
\parbox{1cm}{GRB 980329 (I)}  & \parbox{1cm}{GRB 980329 (II)} & 
\parbox{1cm}{GRB 990123 (I)} & 
\parbox{1cm}{GRB 990123 (II)} 
\\ \hline \hline 
OSSE & Tail & $t_p$ (s) & 3.3(4) & 7.70(2) & 25.16(1) & 38(1)  \\ \hline
& & $\sigma$ (s) & 1.7(4) & 2.62(2) & 2.90(1) & 3(1)  \\ \hline\hline 
& & $t_p$ (s) & 4.33(5) & 6.95(4) & 25.33(1) & 37.08(1)  \\ \hline
& Pulse & $\sigma_r$ (s) & 2.39(5) & 0.74(5) & 3.57(2) & 2.26(2) \\ \hline
& & $\sigma_d$ (s) & 3.6(4) & 4.78(7) & 3.47(2) & 6.07(4)  \\ \hline
& & $\sigma$ (s)   & 3.0(3) & 2.76(6) & 3.52(2) & 4.17(3)  \\ \hline \hline 
BATSE (Ch.\ 3) & Tail & $t_p$ (s) & 3.81(3)& 8.11(1) & 24.63(7)
   & 37.0(4) \\ \hline
& & $\sigma$ (s) & 1.88(4) & 2.35(1) & 2.92(7) & 4.4(4) \\ \hline\hline
& & $t_p$ (s) & 5.18(6) & 10.06(3) & 24.84(3)
   & 36.08(3)    \\ \hline
& Pulse & $\sigma_r$ (s)& 3.37(6) & 2.3(1) &  4.59(4) & 2.85(4)   \\ \hline
& & $\sigma_d$ (s) & 5.28(6) & 1.071(7) & 4.39(5) 
   & 7.48(5)   \\ \hline
& & $\sigma$ (s) & 4.33(6) & 1.69(7) & 4.49(5) & 5.16(5) \\ \hline \hline 
$\Delta$& Tail & $\Delta t_p$ (s) & -0.5(4) & -0.41(2) & 0.53(7) 
       & 1(1) \\ \hline
& & $\Delta\sigma$ (s)    & -0.2(4) & 0.27(2) & -0.02(7) & -1(1)\\ \hline\hline
& Pulse & $\Delta t_p$ (s) & -0.85(8) & -3.11(5) & 0.49(3) & 1.00(3) \\ \hline
& & $\Delta\sigma$ (s) & -1.3(3) & 1.07(9) & -0.97(5) & -0.99(6)\\ \hline\hline
& & $(\Delta t_p)^{{\rm f}}$ (s) & -0.85(8)(35) & -3.11(5)(270) & 0.49(3)(4) 
      & 1.00(3)(0) \\ \hline 
& & $(\Delta \sigma)^{{\rm f}}$ (s) & -1.3(3)(11) & 1.07(9)(80) & -0.97(5)(95)       & -0.99(6)(1)  \\ \hline 
\end{tabular} 
\end{center} 
\caption{{\it As in Table 2, but comparing fits to data 
from OSSE, in the range $1 < E < 5-10$ {\rm MeV},
and Channel 3 of BATSE. }}
\label{table3} 
\end{table}
}

\section{Conclusions and Prospects}

The possibility that the velocity of light might
depend on its frequency, i.e., the corresponding photon energy,
is very speculative. Nevertheless, we consider the motivation
from fundamental physics and the potential significance 
of any possible observation to be sufficient to examine
this possibility in an entirely phenomenological way.
In this paper, we have attempted to extract the maximum
physical information from the few GRBs whose 
cosmological redshifts have been measured, and for which
detailed information on the time distributions of photons in different
energy channels are available.

As could be expected, we have found no significant effect in the data
available, either in the possible delay times of photons of
higher energies, or in the possible stochastic spreads of
velocities of photons with the same energy. If any effect were
to be found in the distributions of photons observed, one
would naturally suspect that it could be due to some source effect.
Therefore, we have made a regression analysis, and found no
significant correlation of either time delays of peak spreading
with the measured redshifts.

We note that, because of the binning (1.024~s) of the BATSE data set
used here. none of the GRBs studied in Tables 2 and 3 exhibits
`microbursts' on timescales $\sim 10^{-2}$~s~(\cite{Scargle}), such as
were
considered previously~(\cite{nature}): if any were to be
discovered in a GRB with known redshift, the sensitivity of the
subsequent analysis would be greatly increased.
We expect that the redshifts of many more GRBs will become
known in the near future, as alerts and follow-up
observations become more effective, for example
after the launch of the HETE~II satellite~(\cite{HETE,Hurley}). Also, it
is clear
that
observations of higher-energy photons from GRBs would be
very valuable, since they would provide a longer
lever arm in the search for energy-dependent effects on photon
propagation. Such higher-energy observations
could be provided by future space experiments such as AMS~(\cite{AMS}) and
GLAST~(\cite{GLAST}).

Even if a correlation with redshift of delays in the arrival times
of energetic photons (or of spreads in their arrival times)
were to be found, one could not
immediately lay the blame on fundamental physics effects on
photon propagation {\it in vacuo}. For example, the GRB
sources might exhibit evolutionary effects that mimic a
correlation with redshift. Alternatively, observational selection effects
might mean that the available GRB samples at low and high redshifts
would have different intrinsic properties, e.g., brightness, that
could also create an artificial correlation with redshift. For
example, there seems to be a general grouping of GRBs into
`short' bursts, with durations $< 2 $~s, and `longer' bursts.
Because of selection effects, 
most of the former are believed to be at small redshifts, so the
analysis could become biased if the sources of `short' and `long' bursts
had different intrinsic time-lags between photons in different
BATSE channels. In point of fact, all the
GRBs whose redshifts have been determined so far are in the `long'
burst category, so this particular problem may not be
important for the data set used here. However, there could
be more subtle selection effects, and any more detailed
phenomenological analysis of the data
should proceed hand-in-hand with more sophisticated
astrophysical modelling of the GRB sources.

Moreover, even if the best efforts of the astrophysical
modellers failed to exclude a fundamental physics effect,
any such
interpretation could be considered established only if other
classes of data were to confirm it. 
This might require analyses
of other types of astrophysical sources, such as AGNs and/or pulsars.
An alternative possibility would be to
consider more carefully laboratory experiments
that might be able to reveal possible variations in the
velocity of light.

\newpage

\section*{Acknowledgements}

\noindent
We thank Giovanni Amelino-Camelia, Marta Felcini, Hans Hofer, 
Shmuel Nussinov, Tsvi Piran, Subir Sarkar and
Spiros Tzamarias for their interest and advice.
The work of N.E.M. is partially supported 
by a P.P.A.R.C. (U.K.) 
Advanced Fellowship, that of V.A.M. 
is partially supported by the 
Greek State Scholarships Foundation, and that of D.V.N. is 
partially supported by DOE grant DE-FG03-95-ER40917.


\begin{thebibliography}{999}


\bibitem[Ahlen {\it et al.}]{AMS}  Ahlen S. {\it et al.}, AMS Collaboration, 
1994, Nucl. Instrum. Meth. A350,
351.


\bibitem[Ashtekar 1999]{ashtekar} 
Ashtekar A., 1999, gr{--}qc/9901023, and references therein.

\bibitem[Amelino-Camelia {\it et al.} 1998]{nature} Amelino-Camelia G., 
Ellis J., Mavromatos N.~E.,
Nanopoulos D.~V. and Sarkar S., 1998, Nature 393, 323;

\bibitem[Amelino-Camelia {\it et al.} 1997]{aemn} 
Amelino-Camelia G., Ellis J., Mavromatos N.~E. and
Nanopoulos D.~V., 1997, Int. J. Mod. Phys. A12, 607. 


\bibitem[Biller {\it et al.} 1998]{biller} Biller S.~D. {\it et. al.}, 
1998, gr-qc/9810044.



\bibitem[Bloom {\it et al.} 1996]{GLAST} Bloom E.~D. {\it et al.}, GLAST
Team, 1996,
{\it Proc. Intern.
Heidelberg Workshop on TeV Gamma-Ray Astrophysics}, eds. H.J. Volk and
F.A. Aharonian (Kluwer, 1996), 109.

\bibitem[Boella {\it et al.} 1997]{beppo} 
Boella G. {\it et al.}, BeppoSAX Project,  1997, Astronomy and
Astrophysics Supplement Series 122, 327.  




\bibitem[Covino {\it et al.} 1999]{polarC} Covino S. {\it et al.},
1999, astro-ph/9906319.

\bibitem[Ellis {\it et al.} 1984]{ehns} 
Ellis J., Hagelin J., Nanopoulos D.~V., and Srednicki M., 1984,
Nucl. Phys. B241, 381.

\bibitem[Ellis, Mavromatos \& Nanopoulos 1992]{emn}  
Ellis J., Mavromatos N.~E., Nanopoulos D.~V., 1992,
Phys. Lett. {B293}, 37. 




\bibitem[Ellis, Mavromatos \& Nanopoulos 1998]{emndbrane} 
Ellis J., Mavromatos N.~E., Nanopoulos D.~V., 1998,
Int. J. Mod. Phys. A13, 1059. 


\bibitem[Ellis, Mavromatos \& Nanopoulos 1999]{emnnew} 
Ellis J., Mavromatos N.~E., Nanopoulos D.~V., 1999,
gr-qc/9904068, Gen. Rel. Grav. in press;
gr-qc/9905048, Gen. Rel. Grav. in press. 


\bibitem[Ellis {\it et al.} 1998]{kanti} 
Ellis J., Kanti P., Mavromatos N.~E., Nanopoulos D.~V. \&
Winstanley E., 1998, Mod. Phys. Lett. A13, 303.  

\bibitem[Ford 1995]{ford}  
Ford L.~H., 1995, Phys. Rev. D51, 1692.





\bibitem[Gambini \& Pullin 1999]{pullin} Gambini R.  and Pullin J., 
1999, Phys. Rev. D59, 124021. 



\bibitem[Garay 1998]{garay} 
Garay L., 1998, Phys. Rev. D58, 124015.  


\bibitem[Hawking, Page \& Pope  1980]{hawking} 
Hawking S., Page D.~N. and Pope C.~N., 1980,
Nucl. Phys. B170[FS1], 283.

\bibitem[Hawking 1982]{hawking2} Hawking S., 1982,
Comm. Math. Phys. 87, 395.
  
\bibitem[Hurley 1998]{Hurley} Hurley K., 1998, astro-ph/9812393.  

\bibitem[Kaaret 1999]{crab} Kaaret P., 1999, astro-ph/9903464.




\bibitem[Landau \& Lifshitz 1975] {landau} Landau L.~D. and 
Lifshitz E.~M., 1975,
{\it Classical Theory of Fields} (Pergamon Press, 1975), Vol. 2, page 257.




\bibitem[Lizzi \& Mavromatos 1997]{lizzi} 
Lizzi F. and Mavromatos N.~E., 1997, Phys. Rev. D55, 7859.



\bibitem[Mavromatos \& Szabo 1998]{ms} Mavromatos N.~E. and Szabo R.~J.,
1998, hep-th/9808124, Phys. Rev. D59, in press. 




\bibitem[Norris {\it et al.} 1996]{norris} Norris J.~R. {\it et al.}, 
1996, The Astrophysical Journal 459, 393. 


\bibitem[OSSE 1999]{osse} OSSE Collaboration, 1999,
{\tt http://www.astro.nwu.edu/astro/osse/bursts/}

\bibitem[Paciesas {\it et al.} 1999]{batse} Paciesas W.~S. {\it et al.},
1999, {\it The fourth BATSE Gamma-Ray-Burst Catalogue} (revised),  
astro-ph/9903205. 

\bibitem[Piran 1999]{Piran} Piran T., 1999, Physics Reports 314,
575.

\bibitem[Rees 1997]{Rees} Rees M., 1997, astro-ph/9701162.

\bibitem[Ricker 1998]{HETE} Ricker G., 1998, in  
{\it Proc. Conference on Gamma-Ray Bursts in the Afterglow Era}, 
Rome, November 1998, Astronomy and
       Astrophysics Supplement Series, to appear;  

\bibitem[Scargle, Norris \& Bonnell 1997]{Scargle} Scargle J.~D.,
Norris J., and Bonnell J., 1997, astro-ph/9712016.

\bibitem[Schaefer 1998]{schaefer} Schaefer B., 1998, astro-ph/9810479. 


\bibitem[Weinberg 1972]{wein} Weinberg S., 1972, {\it Gravitation
and Cosmology: Principles and Applications of the 
General Theory of Relativity} (Wiley, New York, 1972).


\bibitem[Wheeler 1963]{wheeler} 
Wheeler J.~A., 1963, in {\it Relativity, Groups and Topology}, eds. 
B.S. and C.M. de Witt (Gordon and Breach, New York, 1963). 

\bibitem[Wijers {\it et al.} 1999]{polarW} Wijers R.~A.~M.~J. 
{\it et al.},
1999, astro-ph/9906346.

\bibitem[Yu \& Ford 1999]{ford2} Yu H. and Ford L.~H., 1999, gr-qc/9904082. 

\end{thebibliography}
\end{document}